\documentclass[12pt,thmsa]{article}
\usepackage{amssymb}

\input{tcilatex}
\begin{document}

\author{Karl-Georg Schlesinger \\
Institute for Theoretical Physics\\
University of Vienna\\
Boltzmanngasse 5\\
A-1090 Vienna, Austria\\
e-mail: kgschles@esi.ac.at}
\title{The universality question for noncommutative quantum field theory}
\date{}
\maketitle

\begin{abstract}
Present day physics rests on two main pillars:\ General relativity and
quantum field theory. We discuss the deep and at the same time problematic
interplay between these two theories. Based on an argument by Doplicher,
Fredenhagen and Roberts, we propose a possible universality property for
noncommutative quantum field theory in the sense that any theory of quantum
gravity should involve quantum field theories on noncommutative space-times
as a special limit. We propose a mathematical framework to investigate such
a universality property and start the discussion of its mathematical
properties. The question of its connection to string theory could be a
starting point for a new perspective on string theory.
\end{abstract}

\section{Introduction}

The present day understanding of the physical world rests on two fundamental
pillars: First, the structure of space-time which is described by general
relativity (and in this way is directly linked to the gravitational force).
Secondly, quantum field theory as the description of matter and all the
non-gravitational forces of nature. The ongoing task to find a theory of
quantum gravity can be described as the quest for a single consistent
theory, reconciling these two pillars. On the one hand, these two theories
are mutually inconsistent while, on the other hand, there is deep interplay
between them. Let us comment on this interplay in a little bit more detail,
now.

In general relativity one can prove a remarkable theorem about black hole
horizons: In all possible interactions of black holes, the total area of all
black hole horizons, involved, does never decrease. This law can be
formulated mathematically in a way which is completely analogous to the
second law of thermodynamics with the horizon area playing the role of
entropy and the surface gravity of the black hole taking the role of
temperature (see e.g. \cite{Wal}). Indeed, one can give analogous theorems
for all four fundamental laws of thermodynamics. We would like to stress,
here, that these are really \textit{theorems} which can be proved in solely
differential geometric terms. On this level, we have a formal analogy
between two sets of laws with four laws, each. It is tempting to suppose,
then, that this analogy does not come by accident and to interpret the four
laws of black holes as laws for the thermodynamics of black holes. But then
a problem arises:\ We attribute in this way a temperature to a black hole.
But a non-zero temperature of an object in thermodynamics implies a
non-vanishing black body radiation, emitted by this object. Hence, black
holes have to emit radiation! But, by definition of a horizon, black holes
can not emit anything in general relativity. So, there is a contradiction,
here.

At this point, one brings quantum field theory into play. In quantum field
theory, there is a spontaneous production and annihilation of pairs of
particles and antiparticles from the energy of vacuum fluctuations. Now,
imagine such a pair to come into existence very close to the event horizon
of a black hole. Further imagine that during the lifetime of the pair one of
the partners crosses the event horizon. By the laws of general relativity,
this partner will never be back and the pair cannot annihilate any more. In
this way, the partner outside the horizon comes into real existence and is
emitted as radiation (the energy bill for turning a partner in a vacuum
fluctuation into real existence is paid for by energy extraction from the
black hole, i.e. the radiation really appears as emitted from the black
hole). This is, very roughly speaking, the mechanism of Hawking radiation.

What we see from this is that general relativity \textit{needs} quantum
field theory to resolve the contradiction in the thermodynamics
interpretation of some of its fundamental differential geometric theorems.
On the other hand, the two theories are dramatically incommensurable, as can
be seen e.g. from the following observation (see \cite{Wei}):

Suppose, we insert the vacuum energy density resulting from vacuum
fluctuations into the right hand side of the Einstein equation, i.e. into
the energy momentum tensor. It can be easily shown that this leads to an
effective (positive) cosmological constant. Since one does not expect
classical space-time concepts to make sense beyond the Planck length $L$,
one employs a cut-off at $L$ and in this way arrives at a finite value for
the cosmological constant $\Lambda $. Now, a positive cosmological constant
means that light emitted at a space-time point $A$ and absorbed at a
space-time point $B$ acquires a red-shift, just by passing the vacuum.
Assuming that light at $A$ is emitted in the outmost violet part of the
spectrum, we can ask for the distance from $A$ to $B$ such that at $B$ the
light has reached the borderline to the infrared in the spectrum, i.e. one
can calculate a length of maximal visibility for the universe from the
vacuum fluctuations of quantum field theory. The result is that when
standing straight you can hardly see the floor. Hence, general relativity
and quantum field theory fundamentally clash, in this example, by providing
a truly dramatic contradiction to everyday experiment.

The conclusion one has to draw from this seems to be the following: On the
one hand, the deep interplay between the two theories seems to indicate that
general relativity and quantum field theory are viable starting points for
the search for a theory of quantum gravity. On the other hand, it is clear
that these two theories cannot stand just side by side in unmodified form.
This is precisely the perspective, we will take in this paper.

The second fundamental input to this paper comes from the results of \cite
{DFR}. We have already remarked above that one does generally not believe
that a classical description of space-time does hold at or beyond the Planck
length. It has been argued for decades that a foam-like structure of
space-time should appear at this scale. In \cite{DFR} this has been put into
a rigorous form: It is shown there that the creation of black holes from
vacuum fluctuations at the Planck scale leads to uncertainty relations for
space-time coordinates. Sine one knows that uncertainty relations derive
from noncommuting variables in quantum mechanics, this leads to the
conclusion that space-time should be described by a noncommutative space in
a theory of quantum gravity. Accepting the requirement for such uncertainty
relations of space-time coordinates as \textit{universal}, i.e. not linked
to a specific theory of quantum gravity but being a prerequisite of any such
theory, one arrives - together with the considerations from above - at the
following postulate:

\bigskip

\textit{Any theory of quantum gravity should at least have a limit in the
form of a formulation in terms of quantum field theories on noncommutative
space-times}.

\bigskip

We have required a limit, only, since one can imagine that there could exist
theories which incorporate a theory of quantum gravity but go beyond
providing only this. We will call this postulate the \textit{universality
postulate}, in the sequel. The aim of the present paper is to elaborate on
this postulate. First, we try to give a model for a mathematical formulation
of this postulate in section 2. Secondly, in section 3 we undertake to
derive some of the consequences of such a mathematical formulation. As we
will discuss in the next section, this can also be seen as an Ansatz for a
quite different approach to string theory.

We want to stress that this is not a research paper in a proper sense.
Rather it is intended as giving one possible motivation for later technical
work (\cite{Sch} and planed follow-up work) and taking a short view from a
broader perspective.

\bigskip

\begin{remark}
For a paper which also follows the idea of a universality of
noncommutativity but investigates conclusions in a quite different
direction, see \cite{DMK}.
\end{remark}

\bigskip

\section{A mathematical framework}

Consider a smooth $n$-dimensional manifold $M$ endowed with a Poisson
bracket. In \cite{Kon 1997}, it was shown that the deformation theory of
such a manifold is described by a universal formula. Moreover, this formula
derives from a physical system given by a two dimensional field theory (this
description in terms of the Poisson sigma model introduced in \cite{I}, \cite
{SchSt} was made explicit in \cite{CF}).

In analogy, one can consider the following situation: Suppose we would be
given a mathematical description of quantum field theory together with its
deformations to quantum field theories on noncommutative space-times
(henceforth noncommutative quantum field theories or ncQFTs, for short). Of
course, even for the classical case of QFTs on usual Minkowski space a
mathematical formulation is a widely open problem and is considered as one
of the deepest questions of 21st century mathematics. We could introduce,
then, a hypothetical category $\mathcal{A}$ (possibly with some extra
structure, see below) of all (commutative and noncommutative) QFTs. We can
ask, then, if $\mathcal{A}$ is related to some physical system, inheriting
all the symmetry and deformation properties of objects in $\mathcal{A}$
(much the same way as \cite{CF} gives the physical system for deformation
theory of Poisson manifolds). Even more concrete, we can ask if such a
system - if it exists - is related to string- or $M$-theory, in this way, in
principle, opening up the possibility to approach string theory from a
completely different viewpoint by posing the question if it might be related
to a universal deformation theory for quantum field theories (and, hence, be
universal in the sense of the above universality postulate).

Given such a category $\mathcal{A}$, the universality postulate of the
introduction would translate to the postulate that any consistent theory of
quantum gravity should have a formulation in terms of a family of categories 
$\mathcal{B}\left( t\right) $ such that a limit 
\[
\mathcal{B}\left( t\right) \rightarrow _{t\rightarrow 0}\mathcal{A} 
\]
exists. Here, $t$ is a deformation parameter ($t\in \Bbb{R}$) and the family 
$\mathcal{B}\left( t\right) $ and the limit are understood in the sense of
formal deformations.

As we have mentioned, even the formulation of the category $\mathcal{A}$
might be out of reach for a long time to come. We therefore proceed as
follows: We suggest a simpler model for an analogous situation and study
some of its properties. Even in this case, the category $\mathcal{A}$ will
remain a partly conjectural object and one should read the present paper
more as a suggestion for future work rather than a statement of final
results.

\bigskip

Let us pass to the special case of two dimensional conformal field theories.
A central structural element of these theories is the so called operator
product expansion (OPE). The OPE has a rigorous mathematical formulation in
terms of vertex algebras (also called chiral algebras in the more abstract,
coordinate independent, formulation of \cite{BD}). For the case of two
dimensional rational conformal field theories (RCFTs), one can show that the
structure of a vertex algebra completely determines these theories (see \cite
{BPZ}, \cite{Wit}, \cite{FFFS}), i.e. for the highly restricted class of
RCFTs a completely rigorous mathematical formulation is available. One
possible formulation is given in terms of certain inner Frobenius algebras
of braided monoidal categories (see \cite{FS}, \cite{FRS 2001}, \cite{FRS
2003}, \cite{SFR}).

The deformation theory of vertex algebras has been studied in special cases
and shown e.g. to correspond in some examples to massive deformations of
RCFTs (see e.g. \cite{BK}, \cite{FL}, \cite{Fre}, \cite{FR}, \cite{Kli}, 
\cite{Tam}). In an abstract setting (extracting only the general properties
of the product structures in a vertex algebra), we have studied a general
deformation theory of vertex algebras in \cite{GS}. We have shown there that
the deformation theory is described by a system of differential equations
(replacing the Maurer-Cartan equation of the deformation theory of usual
associative algebras) and that this system can formally be derived from an
action principle.

In the next few pages, we will collect some of the properties of the
deformation complex of \cite{GS}. We will suggest, then, a model for the
category $\mathcal{A}$ in terms of these deformation complexes.

\bigskip

We now want to discuss some of the structural properties of the deformation
complex of a quantum vertex algebra as described by the action (18) of \cite
{GS}. We first restrict to the case where, only the structure of one
monoidal category is deformed, i.e. to 
\begin{eqnarray*}
S &\sim &\int \{f\circ d_{\bullet }f+\frac 23f\circ f\circ f+g\circ
d_{\otimes }g+\frac 23g\circ g\circ g \\
&&+\lambda \circ ^2[d_{\otimes }^{\otimes ^2}f-d_{\bullet }^{\otimes
^2}g+f\otimes f-g\bullet g-\left( f\circ ^2g\right) +\left( g\circ ^2f\right)
\\
&&-Comp\left( f,g,g\right) +Comp\left( g,f,f\right) ]\}
\end{eqnarray*}
Remember that on the usual Hochschild complex of associative algebras one
has the structure of a graded homotopy Lie algebra given by the differential
and the Gerstenhaber bracket. Together with the wedge product this combines
into a homotopy Gerstenhaber algebra. In the case of a monoidal category, we
have a more complicated structure on the deformation complex, consisting of
the following elements:

\bigskip

\begin{itemize}
\item  As in the case of the Hochschild complex, we have the composition $%
\circ $, the graded commutator of $\circ $ giving the Gerstenhaber bracket.

\item  We have two different differentials $d_{\bullet }$ and $d_{\otimes }$
plus their liftings $d_{\bullet }^{\otimes ^2}$ and $d_{\otimes }^{\otimes
^2}$to the twofold tensor product.

\item  We have the graded ``curvature'' tensor $Comp$.

\item  The products $\bullet $ and $\otimes $ themselves are represented on
the deformation complex.

\item  Observe that, in contrast to the Hochschild complex, all these
structures are not defined with respect to a single vector space but with
respect to the whole collection of $Hom$-sets.
\end{itemize}

\bigskip

\begin{remark}
The two differentials $d_{\bullet }$ and $d_{\otimes }$ are \textit{not}
compatible in the sense of a double complex. Actually, the compatibility
condition does not make sense on the deformation complex of a monoidal
category because by the composability requirements we have imposed on the
deformations of the composition in \cite{GS}, the two expressions $%
d_{\bullet }d_{\otimes }f$ and $d_{\otimes }d_{\bullet }f$ are not well
defined at the same time.
\end{remark}

\bigskip

The fact that $\bullet $ and $\otimes $ themselves appear on the deformation
complex, again, is a very unusual property. It means that the full structure
of the monoidal category can be reconstructed from the deformation complex.
The monoidal category does not only determine its deformation complex but
also vice versa. This also remains true in the full setting of the action
(18) of \cite{GS}: Besides the structure of both monoidal categories, also
the functor $\mathcal{F}$ appears on the deformation complex and can
therefore be reconstructed from the complex. The following question
therefore arises: Can the additional structure on the complex - appearing
beyond the monoidal category structures of $\mathcal{C}$ and $\mathcal{M}$
and the monoidal functor $\mathcal{F}$ - be interpreted as additional
structure on the triple $\left( \mathcal{M},\mathcal{C},\mathcal{F}\right) $%
? In other words: Can one start with categories $\mathcal{C}$ and $\mathcal{M%
}$ which carry more structure than the one of a monoidal category and a
functor $\mathcal{F}$ adapted to this additional structure such that one
gets a full duality between the triple $\left( \mathcal{M},\mathcal{C},%
\mathcal{F}\right) $ and its deformation complex? If yes, is this additional
structure related to the additional structure which should be carried by a
quantum vertex algebra?

In a first approach toward these questions, we restrict, again, to the
setting of only one monoidal category $\mathcal{C}$ as described by the
action given above. We recall that in \cite{BD} a pseudotensor category is
defined by the following data (where $\mathcal{S}$ denotes the category of
finite sets and surjective maps): A class $\mathcal{M}$ of objects together
with

\begin{enumerate}
\item  ) For any $I\in \mathcal{S}$, an $I$-family of objects $L_i\in 
\mathcal{M}$, $i\in I$, and an object $M\in \mathcal{M}$, there exists the
set $P_I^{\mathcal{M}}\left( \left\{ L_i\right\} ,M\right) $ of $I$%
-operations.

\item  ) For any surjective map $\pi :J\twoheadrightarrow I$ in $\mathcal{S}$%
, families of objects $\left\{ L_i\right\} _{i\in I},\ \left\{ K_j\right\}
_{j\in J}$ and an object $M$, there exists a composition map 
\[
P_I^{\mathcal{M}}\left( \left\{ L_i\right\} ,M\right) \times \prod_{i\in
I}P_{J_i}^{\mathcal{M}}\left( \left\{ K_j\right\} ,L_i\right) \rightarrow
P_J^{\mathcal{M}}\left( \left\{ K_j\right\} ,M\right) 
\]
with 
\[
\left( \varphi ,\left( \psi _i\right) \right) \longmapsto \varphi \left(
\psi _i\right) 
\]
and $j\in J_i$ in $P_{J_i}^{\mathcal{M}}\left( \left\{ K_j\right\}
,L_i\right) $ where $J_i=\pi ^{-1}\left( i\right) $.
\end{enumerate}

\bigskip

The composition is assumed to be associative and the existence of units is
assumed (see \cite{BD}). Recall that a pseudotensor category with one object
is an operad. A representable pseudotensor category is equivalent to a usual
category $\mathcal{M}$ together with functors 
\[
\otimes _I:\mathcal{M}^I\rightarrow \mathcal{M} 
\]
for any $I\in \mathcal{S}$ and a natural compatibility morphism 
\[
\epsilon _\pi :\otimes _JK_j\rightarrow \otimes _I\left( \otimes
_{J_i}K_j\right) 
\]
- for any surjective map $\pi :J\twoheadrightarrow I$ - satisfying certain
naturality conditions (see \cite{BD}). If all $\epsilon _\pi $ are
isomorphisms, the pseudotensor category reduces to a symmetric monoidal
category.

Now, observe that the deformation complex of a monoidal category $\mathcal{C}
$ is given by the multilinear maps 
\begin{equation}
Morph\left( \mathcal{C}^k\right) \rightarrow Morph\left( \mathcal{C}\right)
\label{1}
\end{equation}
for any $k\in \Bbb{N}$ (where the multilinear maps are understood as those
which are multilinear locally on the $Hom$-sets). Considering the set of
such multilinear maps for fixed $k$ as the set of $k$-operations, one can
define a composition on the deformation complex in complete analogy to the
composition in a pseudotensor category. One has a kind of pseudotensor
2-category with one object in this way (i.e. a kind of categorical operad).
We write ``a kind of'' because on the one hand, all the maps in (\ref{1})
live on the category $\mathcal{C}$ but on the other these maps are \textit{%
not} functors. If we want to stress this fact, we will speak of a \textit{%
pseudotensor semi-2-category}.

It is a straightforward consequence that the tensor $Comp$ is induced from
this pseudotensor composition as a special case. So, we get a kind of
categorical version of the pseudotensor structure of \cite{BD} as part of
the additional structure on the monoidal category $\mathcal{C}$, given by
the deformation complex.

We phrase the following question:

\bigskip

\textbf{Question:} Can the full additional structure, given by the
deformation complex, be understood as a (kind of categorical) extension of
the chiral algebra structure of \cite{BD} (possibly after extending the
consideration to the full triple $\left( \mathcal{M},\mathcal{C},\mathcal{F}%
\right) $)? Does a duality between the triple $\left( \mathcal{M},\mathcal{C}%
,\mathcal{F}\right) $ and its deformation complex hold?

\bigskip

Suppose the answer to the above question would be in the affirmative. The
general noncommutative deformation of a RCFT could then be understood as
being given by what a category theorist would probably call a 2-chiral
algebra. In the sequel, let us assume this to be true. Whatever the precise
definition of a 2-chiral algebra will be (we assume it to resemble the
structure of the deformation complex, described above), a 2-chiral algebra
will certainly be given by a bicategory with some extra structure. It is
this lack of a precise definition of a 2-chiral algebra and our ignorance of
the answer to the above question which makes even the simple toy model we
suggest for $\mathcal{A}$ a largely speculative object. Assuming that we
have bicategories with some extra structure, all 2-chiral algebras should
ensemble into a tricategory (see \cite{GPS} for the definition of
tricategories or the introduction included in \cite{Lau}). This tricategory
will be our toy model for $\mathcal{A}$.

\bigskip

\section{Some properties of $\mathcal{A}$}

Let us discuss, now, some of the possible properties of $\mathcal{A}$. This
discussion will, again, be largely speculative. Remember that we understand
this paper, only, as a motivation for later technical work.

In formal deformation theory, one can often distinguish between the
deformations of an algebraic structure inside the given category and the -
more general - quantum deformations which usually lead to a categorification
of the structure. E.g. one has deformations of Lie algebras - or their
universal envelopes - as such and quantum deformations inside the more
general category of Hopf algebras. In \cite{Sch 2001}, we have given an
argument that certain monoidal bicategories should not have nontrivial
quantum deformations, i.e. all deformations should remain in the given
category of these special monoidal bicategories. We called this
ultrarigidity.

Let us speculatively assume that ultrarigidity also holds true for 2-chiral
algebras (observe that the structure on the deformation complex, described
in the previous section, seems to be a generalization of a class of monoidal
bicategories). It follows, then, that all deformations of 2-chiral algebras
remain inside $\mathcal{A}$. If there would be a deformation of a 2-chiral
algebra to an algebraic structure outside $\mathcal{A}$, this should
correspond to a deformation of $\mathcal{A}$ itself. Conversely, a
deformation of $\mathcal{A}$ should correspond to a deformation of the
notion of 2-chiral algebra, i.e to a deformation of the objects of $\mathcal{%
A}$. If ultrarigidity holds for 2-chiral algebras, this implies that any
deformation of $\mathcal{A}$ should correspond to a functor $\mathcal{F}$ of
tricategories, 
\begin{equation}
\mathcal{F}:\mathcal{A}\rightarrow \mathcal{A}
\end{equation}
(If it would not be a true functor of tricategories, this would mean that
the morphism structure of $\mathcal{A}$ would be changed in the process of
deformation. But this, in turn, would mean that the notion of 2-chiral
algebra has been deformed, i.e we would have deformations of 2-chiral
algebras which do not stay inside $\mathcal{A}$, contradicting
ultrarigidity. Hence, $\mathcal{F}$ has to be a functor of tricategories,
once we have assumed ultrarigidity.). Conversely, if any deformation of $%
\mathcal{A}$ would lead to a functor (\ref{1}), this should imply
ultrarigidity. Observe that $\mathcal{F}$ does not have to be invertible or
even an equivalence of tricategories. So, there are still two possibilities
for deformations: On the one hand, the trivial deformations corresponding to
equivalences of $\mathcal{A}$ and, on the other hand, the nontrivial
deformations which can e.g. lead to proper sub-tricategories of $\mathcal{A}$%
. So, there is still room for quantization of 2-chiral algebras in the sense
of nontrivial deformations of $\mathcal{A}$ but these would be 2-chiral
algebras, again. Especially, ultrarigidity would imply that quantization of
2-chiral algebras has to be functorial.

So, assuming that - in the narrow sense of our toy model - commutative and
noncommutative quantum field theories would be described by 2-chiral
algebras, all the important questions how such quantum field theories
ensemble into families, i.e. how a single quantum field theory is deformed
in a family, how they are renormalized (as a special case of deformation
theory in a family of quantum field theories), etc. should have a
counterpart in the deformation theory of $\mathcal{A}$. Especially, with a
detailed knowledge about the correct definition of 2-chiral algebras and the
tricategory $\mathcal{A}$, it should be possible to prove ultrarigidity for
2-chiral algebras from the deformation theory of $\mathcal{A}$ by checking
for the existence of the functors (\ref{1}). As a very first step, we will
study the deformation theory of tricategories in the forthcoming paper \cite
{Sch}.

\bigskip

\begin{remark}
Assuming the universality postulate from above, proving ultrarigidity from
the deformation theory of $\mathcal{A}$ would imply that $\mathcal{A}$ is
singled out as providing the only consistent framework for a quantum theory
of gravity. A possible physical system related to $\mathcal{A}$ (see below)
would in this way be singled out on the basis of the universality posulate
and mathematical - deformation theoretic - arguments.
\end{remark}

\bigskip

There are two important questions, then:

\bigskip

\begin{itemize}
\item  Does the deformation theory of $\mathcal{A}$ fully determine the
structure of $\mathcal{A}$, much the same way we have discussed this for the
deformation theory of vertex algebras, in the previous section?

\item  Does the deformation theory of $\mathcal{A}$ (deformation equations,
replacing the Maurer-Cartan equation, cohomology, etc.) arise from a
physical system?
\end{itemize}

\bigskip

Since chiral algebras correspond to two dimensional quantum field theories,
one would suspect - in the spirit of the dimensional ladder and
categorification - that 2-chiral algebras should correspond to field
theories on the three dimensional world volume of a membrane (possibly
topological, since usual chiral algebras correspond to RCFTs). If this would
be true, it would naturally fit in with the idea of \cite{Str} that
noncommutative quantum field theories should, in a very general sense, arise
from open membrane systems. So, it seems natural to sharpen the second of
the above questions to the question if the deformation theory of $\mathcal{A}
$ could arise from (topological)\ $M$-theory. If the answer would be in the
affirmative, this would open up the possibility to study $M$-theory as the
universal theory related to the deformation theory of (noncommutative)
quantum field theories (much the same way as \cite{CF} gives the universal
theory related to the deformation theory of Poisson manifolds). It will need
rigorous technical work to judge if such a suggestion is anything more than
pure speculation.

\bigskip

\begin{remark}
For a different approach which also suggests that string theory might have
universality properties as a theory ruling the behaviour of QFTs, see \cite
{Vaf}.
\end{remark}

\bigskip

\textbf{Acknowledgments:} I would like to thank H. Grosse and V. Mathai for
discussions on or related to the material of this paper.

\end{document}